\documentstyle[11pt,rafringe,twoside,epsf]{article}

\def\arg#1{{\it#1\/}}
\let\prog=\arg

\def\edcomment#1{\iffalse\marginpar{\raggedright\sl#1\/}\else\relax\fi}
\reversemarginpar

\begin{document}
\title{What can we learn about extragalactic radio jets from X-ray data?}
 \author{D. E. Harris}
\affil{SAO, 60 Garden St., Cambridge, MA 02138, USA; harris@cfa.harvard.edu}

\begin{abstract}
We review the current status of resolved X-ray emission associated
with extragalactic radio jets and hotspots.  The primary question for
any particular jet is to decide if the X-rays come from the
synchrotron process or from inverse Compton scattering.  There is
considerable evidence supporting synchrotron emission for knots in the
jets of FRI galaxies.  For FRII terminal hotspots detected in the
X-ray band, synchrotron self-Compton emission continues to provide
viable models with one possible exception (so far).  Inverse Compton
scattering on photons of the cosmic microwave background is indicated
for a few powerful jets, and is expected to be an important
contributor if not the dominating mechanism for higher redshift
objects.  The application of a model generally yields physical
parameters and in many cases, these include the Doppler boosting
factor. 
\end{abstract}

\section{Introduction}


The study of relativistic jets via X-ray emission provides certain key
advantages unobtainable at lower frequencies.  If the X-rays come from
synchrotron emission, then we are dealing with Lorentz energies
$\gamma\approx10^7$.  This fact leads to two important features: $E^2$
halflives are very short which means that observed emission regions
correspond to sites of injection or acceleration of particles, and
that variability timescales are of order a year or less.  For the most
part, X-ray variability has been limited to unresolved cores; whereas
now we can observe variability for knots which are well separated from
the nucleus (Harris et al. 2003).

If the X-ray emission arises from the inverse Compton (IC) process, 
then we are dealing with electrons for which $\gamma\leq1000$ and
thus we can obtain amplitudes of the electron spectra at energies not
available to ground based radio observations.

For the case of synchrotron self-Compton emission (SSC), we are able
to estimate the average magnetic field strength and for the case of
equipartition fields, constrain the filling factor and the
contributions to the total particle energy density from low energy
electrons and from protons.

In this overview which is limited to jet knots and hotspots clearly
distinct from emission associated with the nucleus, we discuss the
emission processes and summarize what we have learned so far.  We end
with a discussion of some of the current problems and goals.  We
concentrate on aspects for which new information has become available
subsequent to a previous short review prepared for a meeting in 2000
August (Harris 2002) and discussions on the enhancement of IC
scattering on the cosmic microwave background (CMB) experienced by
relativistic jets (Celotti, Ghisellini, \& Chiaberge. 2001; Tavechhio
et al. 2000; Harris \& Krawczynski 2002).

We maintain a website (http://hea-www.harvard.edu/XJET/) which
currently lists 37 radio galaxies and quasars with known X-ray
emission from jets or hotspots.  Before the launch of the Chandra
X-ray Observatory, this table had only 7 entries: 3C273, Cen A, M87,
Cyg A, Pic A, 3C120, and 3C390.3.

\section{X-ray Emission Processes and what they reveal}

For any given feature, it is necessary to identify the emission
process before we can estimate physical parameters.  For knots and
hotspots there is no longer any serious consideration given to thermal
emission, although X-ray emission from hot gas may be closely
associated with jets in some cases.  The main problem is to
decide between synchrotron and IC emissions (the latter comes of
course in many different flavors depending on the nature of the target
photons).  Every synchrotron source must also produce IC emission from
at least the CMB and the synchrotron photons; our
question is, which process dominates the X-ray emission.

\subsection{Synchrotron Self-Compton Emission}

SSC emission models have been successfully applied to several terminal
hotspots of FRII radio galaxies and the resulting magnetic field
strengths are close to equipartition for filling factors close to 1 and
little or no contribution to the particle energy density from protons
(Harris 2002).  Hardcastle et al. (2002) have recently added to the
small number of FRII hotspots with sufficiently large photon energy
densities to provide detectable SSC X-rays.  In figure 1 we
show one of their sources, the double hotspot in the quasar 3C~351.
This is not a typical SSC hotspot, nor is it a typical FRII radio
structure since the hotspot on the other side of the source is very
weak.  It is one of the few where the X-ray emission is substantially
greater than predicted by SSC with equipartition fields.  Here, and in
the western hotspot of Pic A, Doppler boosting, a significant
departure from equipartition, and/or a substantial contribution from
synchrotron emission are possible explanations for the excess
emission.

\begin{figure}\label{fig:hs}
\plotfiddle{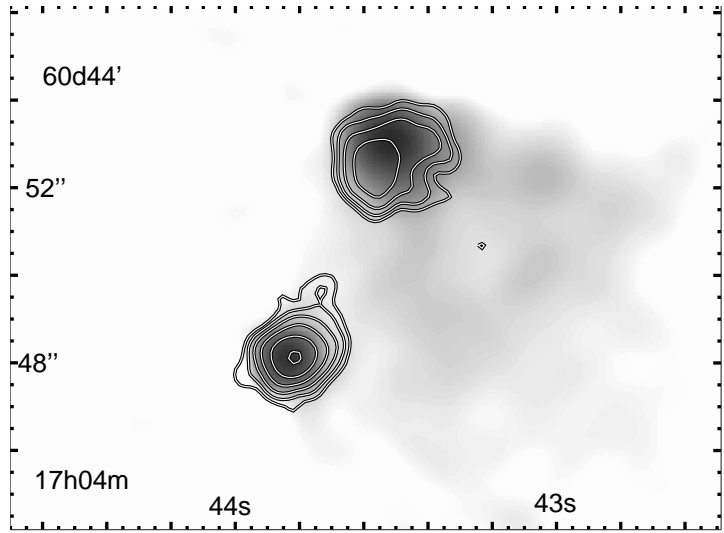}{2.45in}{0}{50}{50}{-150}{-100}
\caption{The double hotspot system in 3C351.  The greyscale shows an
8GHz VLA map with a beamsize of 0.86$''$, comparable to the Chandra
resolution.  The contours are from the Chandra X-ray image smoothed
with a Gaussian of FWHM=$0.5''$.  Contour levels increase by factors
of two.  Note that the northern hotspot is well resolved in the
X-rays, and that there is a significant mismatch between the X-ray and
radio morphologies.  For most SSC hotspots, the X-ray morphology
closely mimics the radio structure.}
\end{figure}

\subsection{IC/CMB}

The notion that kpc scale jets of FRII sources are still moving with
significant bulk relativistic velocity was argued by Bridle
(1996) and others.  Celotti et al. (2001) and Tavecchio et al. (2000)
suggested that in the jet frame, the photon energy density, $u(\nu)$,
of the CMB would be augmented by $\Gamma^2$ ($\Gamma$ is the jet's
Lorentz factor), and this could shift the
primary energy loss mechanism from the normally dominate synchrotron
channel to the IC channel, thus increasing the ratio of the IC to
synchrotron luminosities.  These authors successfully applied this
idea to the jet of PKS0637 and derived $\Gamma\approx~10$ for
equipartition fields.  However, when the model is applied to low
redshift FRI radio galaxy jets, it fails since extremely large values
of $\Gamma$, coupled with unreasonably small angles of the jet to the
line of sight, are required (Harris \& Krawczynski 2002).

Once the redshift becomes significant, the (1+z)$^4$ term in the
energy density of the CMB becomes increasingly important, and for such
sources, the 'heavy lifting' to increase u($\nu$) comes from the
redshift rather than the $\Gamma^2$ term (see e.g. Siemiginowska et
al. 2002).

\subsection{Synchrotron Emission}

Synchrotron X-ray emission is normally modeled by extending the
powerlaw distribution of electron energy (with or without breaks) to
Lorentz factors, $\gamma$, of $10^7$ to $10^8$.  This sort of model
can be applied to most or all X-ray knots in the jets of FRI sources and
does not require excessive amounts of energy.  It is true however
that the E$^2$ lifetimes are short, and hence observed X-ray emission
must demarcate injection/acceleration regions.

When constructing synchrotron (or IC) models, we often obtain
estimates of the beaming parameters as a by product.  In the case of
the knot HST-1 in the M87 jet, we were able to demonstrate that the
variability characteristics were consistent with a synchrotron model,
but not with IC emission (Harris et al. 2003).  Moreover, by solving
for synchrotron parameters for a range of Doppler factors $\delta$, it
was found that the observed decay rates in the lightcurves (harder
X-rays decay faster than softer X-rays) were consistent with
synchrotron losses in a field of order 1 mG with $\delta\approx~3$.

To relate the decay of the lightcurve to a synchrotron loss model, we
start with the normal synchrotron equations (Pacholczyk 1970) and make
the simplifying assumption that if the emission in a small band drops
by a certain amount, then there will be a proportionally smaller
number of electrons after the drop.  Although this approach assumes
that the magnetic field strength remains constant, it demonstrates the
method of obtaining parameters.

The electron spectrum is the usual power law.

\begin{equation}
N(E)~=~k_e~E^{-p}
\end{equation}

Synchrotron losses go as $E^2$ ($c_1$ \& $c_2$ are constants):

\begin{equation}
\frac{dE}{dt}~=~c_2~B^2~E^2~erg~s^{-1}
\end{equation}

Consider a segment of the electron energy distribution responsible for
a given observed X-ray band between $\nu_1$ and $\nu_2$.  Let
$h\equiv\frac{\nu_2}{\nu_1}$.  The number of electrons entering the
band (per sec)
at $E_2$ will be $N(E_2)\times~\frac{dE_2}{dt}$ and the number leaving
at $E_1$ will be $N(E_1)\times~\frac{dE_1}{dt}$.

The net change will be\\

\begin{equation}
\Delta~N=N(E_1)~\frac{dE_1}{dt}~[h^{1-\frac{p}{2}}~-~1]~s^{-1}
\end{equation}

and for a drop in flux, f(t), over some given time, $t'=t'_2~-~t'_1$ (primes
denote the jet frame):

\begin{equation}
\frac{f(t'_2)}{f(t'_1)}= \frac{\int~N(E,t'_1)~+~t'\times\Delta~N}{\int~N(E,t'_1)}
\end{equation}

With $t'=\delta~t_o$ and $E'=\sqrt{\frac{\nu}{\delta~c_1~B}}$, we
find:

\begin{equation}
\delta~B^3~=~\frac{c_1}{\nu_1}~\frac{1}{t_o^2~c_2^2}~\left[1~-~\frac{f(t_2)}{f(t_1)}\right]^2~~\frac{1}{(p-1)^2}~\left[\frac{h^{(1-p)/2}-1}{h^{(1-p/2)}-1}\right]^2
\end{equation}

All the terms on the far right depend only on the spectral index and
the bandwidth.  In the center we have the ratio of fluxes at two
times, and the time between intensity observations is given by t$_o$.
Thus on the right are observables, providing a measure of $\delta~B^3$
which can be compared with this product from trial $\delta$s and
resulting equipartition fields (see Harris et al. 2003 for the
relevant table).  For the 3 energy bands used in the M87 analysis of
HST-1, we find $\delta=3.3\pm1$ (slightly different values are
obtained if $\alpha\neq1.0$)

This 'modest beaming' model is consistent with beaming parameters
deduced from proper motions of substructures in the same knot at optical
wavelengths (Biretta, Sparks, \& Macchetto 1999).


\section{Problems}

\subsection{High frequency, flat spectrum components - the 'bowtie'
problem:  Some knots have flatter X-ray spectra than they 'should'.}

It has become common to differentiate IC from synchrotron emissions on
the basis of the X-ray spectral index.  If $\alpha_x\leq\alpha_r$,
then IC emission is indicated since we expect the low frequency radio
spectrum to extrapolate to unobservable frequencies with $\alpha_r$ or
less.  If $\alpha_x\geq\alpha_r$, then losses or cutoffs in the
synchrotron spectrum are invoked.  We believe that this approach
is reasonable, but should not be considered as definitive.  Fresh
injection may be manifest by flat spectra at high energies, and at the
other end of the electron spectrum, we don't actually know what slope
the power law distribution (if indeed it is still a power law) might
have at energies below those responsible for synchrotron radiation
observable from the ground.

The 'bowtie problem' arises from the expectation that synchrotron
spectra (log S$_{\nu}$ vs. log $\nu$) are generally concave downwards
and such a spectrum extrapolated from the optical data cannot
accommodate the 'bowtie' delineating the X-ray intensity and spectral
index (with uncertainties) for some jet knots and hotspots; see e.g.  Pic A
(Wilson, Young, \& Shopbell 2001) or M87 (Wilson \& Yang 2002).
Although we have argued in the past for 'concave downwards' sort of
spectral fits, we no longer believe that this argument is germane to
knots in jets.  Rather it is a characteristic of the continuous
injection (CI) model for stationary sources.  For relativistic jets,
the electrons resulting from a power law generation at a shock are
convected downstream instead of accumulating in the local emitting
volume as assumed for CI.  Our view is that while synchrotron spectra
of jet knots may be a single or broken power law over a wide frequency
range, at acceleration sites (wherever X-rays are observed), it is
expected that there may be a flat spectrum component visible only at
the highest frequencies because the lower segment of such a power law
is lost in comparison to the much higher intensity component which has
arrived from upstream and no longer contains electrons with
$\gamma\approx10^7$.  The only non-standard ingredient of this
scenario is that the local acceleration site should not intercept all
of the flow from upstream, thus allowing, in a sense, two spectral
components: the upstream contribution which is cutoff at a lower
energy and the locally (re)accelerated flat spectrum component.

\subsection{Offsets in peak brightnesses between bands \& 
progression of relative intensities in different bands}

Even when care has been exercised to ensure that the effective
beamsizes are close to being the same, relative offsets are sometimes
observed between peak emissions at different bands.  In fig.~2, we
show a comparison of radio and X-ray images of M87.  Offsets can be
seen for knots D, E, and F (2.5$''$ to 9$''$ from the core).  Other
cases are the nearest X-ray jet (Cen A, Kraft et al. 2002) and one of
the more distant jets, the quasar PKS1127-148 (z=1.18; Siemiginowska
et al. 2002).

\begin{figure}\label{fig:m87}
\plotone{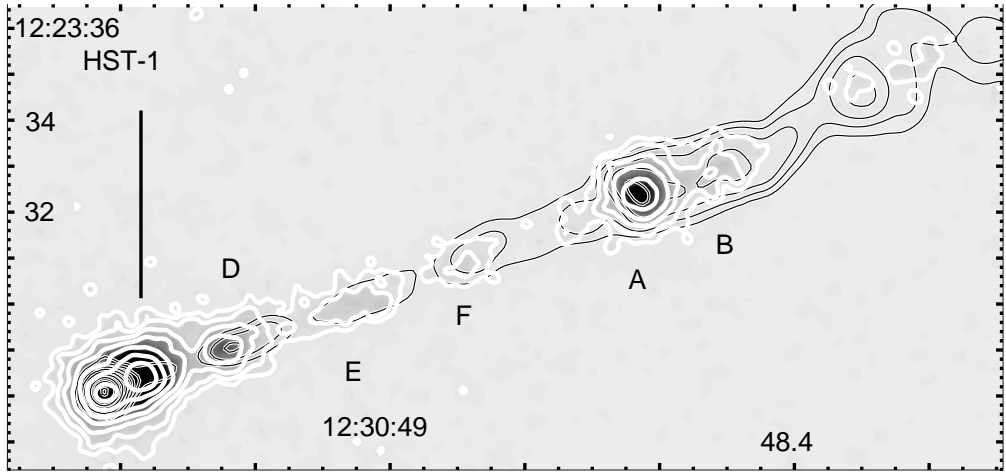}
\caption{Radio and X-ray images of M87.  The greyscale and fat white
contours are the Chandra data with 0.5$''$ resolution (FWHM).  Contour
levels increase by factors of two and the lowest contour is
1.0$\times10^{-16}$~ergs cm$^{-2}$ s$^{-1}$ per 0.049$''$ pixel in the
0.2 to 6 keV band.  The thinner contours (combination of black and
white) show the 8GHz VLA map with a beamsize of 0.3$''$, comparable to
the Chandra resolution.  Note the offsets between the radio and X-ray
brightnesses for knots D, E, \& F.}
\end{figure}

For synchrotron models of FRI radio galaxies (i.e. with physical
offsets of order parsecs, not kpc), a reasonable explanation of the
offsets would be that the acceleration region is distributed along the
jet and the magnetic field strength is increasing after the initial
shock front.  Moving downstream, the increasing field strength will
lower the maximum energy attainable in subsequent shocks and increase
the emissivity.  Since the halflives of X-ray emitting electrons are
of order a year, whereas the optical and radio emitting electrons
would endure ten and 10,000 times longer (respectively), such an
acceleration region would naturally produce the observed offsets.  No
offsets would be expected for a single shock, although the lower
frequency emissions could extend further down the jet than the X-ray
knot.

A somewhat similar situation exists for the overall structure of a few
jets: the brightest X-ray features are close to the nucleus and the
knots get fainter moving out the jet.  In the case of 3C~273, the
optical features have similar intensities but the radio intensities
increase as one moves further downstream (the final radio 'knot' is
weak or undetected at optical and X-ray energies).  If one were to
observe the 3C~273 jet from a great distance with only a few
resolution elements for the jet, one would see an offset between the
radio and X-ray peaks.  Put another way, each small segment of some
jets mimics the gross characteristics of the whole (observable) jet.
Once again, a natural explanation would be an increasing average field
strength as one moves out along the jet.

\subsection{Other items needing attention}

\begin{itemize}

\item{Is there any method to check on the extrapolation to lower
$\gamma$ in the electron spectra which is required for IC/CMB models?}

\item{Are there serious departures from B$_{eq}$?}

\item{Is complex jet structure required?  Celotti et al. (2001)
suggested the possibility of a fast spine with large $\Gamma$,
surrounded by a sheath with low $\Gamma$.  This obviously provides
more latitude to explain observed fluxes in different bands, but is it
necessary?}

\item{How much beaming is there, and where (i.e. in FRI jets? in
hotspots?)?  Currently we think $\Gamma$ is of order a few in FRI
jets, but we don't really have convincing limits for FRII jets.}

\item{X-ray emission between discrete knots.  If, as seems likely in
several jets, there is quasi continuous X-ray emission along the jets,
then the standard synchrotron model of acceleration at discrete
locations (shocks=knots) fails to explain the emission between knots.
Two possibilities are IC/CMB emission from cold pairs (Harris \&
Krawczynski 2002) and synchrotron emission from distributed
acceleration such as might occur from magnetic reconnection along a
magnetically dominated jet (Blandford, private communication).}

\end{itemize}

\section {Summary}

\subsection{Goals}
It seems reasonable that we may expect progress in understanding a
number of jet properties in the not too distant future.

\begin{itemize}

\item{Composition: is the major carrier of momentum Ponyting flux,
normal plasma, or pair plasma?  NB: large $\Gamma$s at large distances
(kpc scales) effectively kills hot pairs.}

\item{Distribution of $\Gamma$: refining models, we should be able 
to obtain reasonable estimates of $\Gamma$ for various jets and for
features within jets.}

\item{Jet Structure: does the fluid follow a helical path or is there
a spine plus sheath structure?}

\end{itemize}

\subsection {What have we learned so far?}

Except for terminal hotspots, all detected X-ray jets are one sided.
Therefore, whatever the emission process, Doppler favoritism is
operating and we need to include beaming effects.  With increased
sensitivity, examples of bona fide two sided jets will most likely be
found, but this is expected for the small values of the bulk
relativistic velocity currently estimated for many FRI radio galaxies.

As reviewed in Harris (2002), X-ray emission from terminal hotspots
provides the following conclusions.  If the X-rays are synchrotron
emission from high energy electrons which are generated by Proton
Induced Cascades (rather than the usual high energy tail of a power
law distribution), the average magnetic field strength will be large
(of order a mG) and hotspots would be good candidates for the origin
of UHE cosmic rays.

If the hotspot X-rays come from SSC emission as is commonly accepted,
then the fact that the average magnetic field required for SSC is
consistent with the conventional equipartition field strength, can be
taken as circumstantial evidence that relativistic protons do not
contribute to the total particle energy density, and thus are most
likely absent.

For the handful of hotspots successfully modeled with SSC, we find
$\alpha_x~\approx~\alpha_r$, as expected,  and the ratio of photon to
magnetic energy densities $\leq~0.1$

From X-ray synchrotron emission from knots in jets,
$\gamma~\approx~10^7$ and the halflife of the highest energy
electrons, $\tau_{\frac{1}{2}}\leq$~a few years.
Therefore, X-rays mark the spot of acceleration sites.

From models involving IC/CMB emission, we find the following.

\begin{itemize}

\item{Since $\nu_{IC/CMB} =
3\times10^{11}\times\Gamma\times(1+z)\times\gamma^2$, the electrons
responsible for the observed X-rays have energies $\gamma$=30-300 and
we may obtain an estimate of the amplitude of the electron spectrum at
very low energies.}

\item{Some jets are relativistic on kpc scales.}

\item{$\alpha_x~\approx~\alpha_r$ (or flatter).}

\end{itemize}

\acknowledgments

The radio map of M87 was kindly supplied by F. Owen.
The work at SAO was supported by NASA contract NAS8-39073 and
grant GO2-3144X.


\end{document}